# Electrical control of lifetime-limited quantum emitters using 2D materials

K. G. Schädler[1], C. Ciancico[1], S. Pazzagli[2,3], P. Lombardi[2]

A. Bachtold[1], C. Toninelli[2,4], A. Reserbat-Plantey[1] and F.H.L. Koppens[1,5]

[1]ICFO-Institut de Ciencies Fotoniques, The Barcelona Institute of Science and Technology, 08860 Castelldefels, Spain

[2]LENS and CNR-INO, Via Nello Carrara 1, 50019 Sesto Fiorentino, Italy

[3]Dipartimento di Fisica ed Astronomia, Università di Firenze, Via Sansone 1, 50019 Sesto Fiorentino, Italy

[4]QSTAR, Largo Fermi 2, 50125 Firenze, Italy

[5]ICREA – Institució Catalana de Recerça i Estudis Avancats, 08010 Barcelona, Spain.

**Solid state quantum emitters are a mainstay of quantum nanophotonics as integrated single photon sources (SPS) and optical nanoprobes. Integrating such emitters with active nanophotonic elements is desirable in order to attain efficient control of their optical properties but typically degrades the photostability of the emitter itself. Here, we demonstrate a tuneable hybrid device that integrates lifetime-limited single emitters (linewidth $\sim 40$ MHz) and 2D materials at sub-wavelength separation without degradation of the emission properties. Our device's nanoscale dimensions enable ultra-broadband tuning (tuning range > 400 GHz) and fast modulation (frequency $\sim 100$ MHz) of the emission energy, which renders it an integrated, ultra-compact tuneable SPS. Conversely, this offers a novel approach to optical sensing of 2D material properties using a single emitter as a nanoprobe.**

Hybrid nanophotonic systems blend the strengths of distinct photonic elements to strongly enhance light-matter interactions[1] in integrated photonic circuits. In these systems, narrow-linewidth quantum light emitters play a key role as single photon sources (SPS) which interact with their nanoscale environment[2,3]. Controlling these interactions provides versatile SPS tuning[4] required for coupling quantum resources[5–7]. Integrating nanoscale light emitters with two-dimensional (2D) materials is motivated by the rich physics of near-field interactions[8] and new hybrid light-matter states[9,10]. This approach unites integrated solid-state SPS such as nitrogen vacancy centres[11], quantum dots[12] and single molecules[13] with the diverse optoelectronic properties of 2D materials that facilitate emitting[14], controlling[15–17] and detecting[18] light at the nanoscale. In such hybrid devices, quantum emitters can be integrated at sub-wavelength separation to the 2D interface to achieve efficient near-field coupling[8], which modifies the emitter's radiative decay rate[19–21] or transition energy[22,23]. Recent experimental studies integrated 2D materials with ensembles of broadband emitters to demonstrate electrical[24–26] and electromechanical[27] tuning of the decay rate by controlling non-radiative energy transfer (nRET) or the energy flow to confined electromagnetic modes such as 2D polaritons[26,28]. Therefore, hybrids of 2D materials and SPS have the potential for *in situ* control of the conversion and channelling of single photons at the nanoscale. So far, these studies have been limited to ensembles and broad linewidth emitters. Integrating bright and narrow quantum emitters in such systems paves the way towards a tuneable quantum light-matter interface, which is an essential ingredient for integrated quantum networks.

Here, we demonstrate hybrid integration of 2D materials (semi-metallic graphene or semi-conducting $MoS_2$) with single, lifetime-limited quantum emitters in nanocrystals to provide active emission control. Using the 2D materials as transparent electrodes, we show broadband Stark tuning of the emission energy over 40,000 times the emitter linewidth and fast modulation of the emitter's optical resonance on the time scale of its radiative lifetime. Such tuning can mitigate inhomogeneous broadening in solid-state environments to enable resonant and synchronised interaction between distinct quantum systems[5] and allows for controlled coupling of narrowband quantum emitters to broadband nanophotonic circuitry[29]. Our approach is particularly suited for making integrated devices: although the 2D material is just tens of nanometres from the quantum emitter, we observe only weak emission linewidth broadening and spectral diffusion. In contrast, emitters close to bulk transparent electrodes[30] such as ultra-thin metal films suffer strong emission quenching[31], which hampers nanoscale integration. Furthermore, we find that the deposition process required to integrate a transparent conducting oxide inhibits molecular fluorescence, which highlights the potential of 2D materials for integration with ultra-sensitive quantum emitters. At the same time, we show that a single quantum emitter can be used as a transducer of the 2D materials' electronic properties.

We choose single dibenzoterrylene (DBT) molecules as bright, photostable single photon sources[32,33] emitting at 785 nm (1.58 eV) with lifetime-limited linewidth[34] (~ 40 MHz) at 3 K even when hosted in a sub-micron environment[35]. Experimentally, we perform scanning laser spectroscopy to address individual DBT molecules at sub-wavelength separation to a 2D metallic or semiconducting interface. The device comprises DBT molecules embedded in anthracene nanocrystals (see Supplementary Note 1), interspersed in a thin polyvinyl alcohol (PVA) film on a $Si^{++}/SiO_2$ substrate (Fig. 1a). $MoS_2$ or graphene flakes are placed on top of the polymer film by a dry transfer technique and electrically contacted by gold electrodes. This transfer process is performed at room temperature and without other nanofabrication steps which potentially degrade emission[36]. Spatial maps of DBT ensemble fluorescence (see methods) at 3 K show bright, localised emission beneath both 2D materials (Fig. 1b).

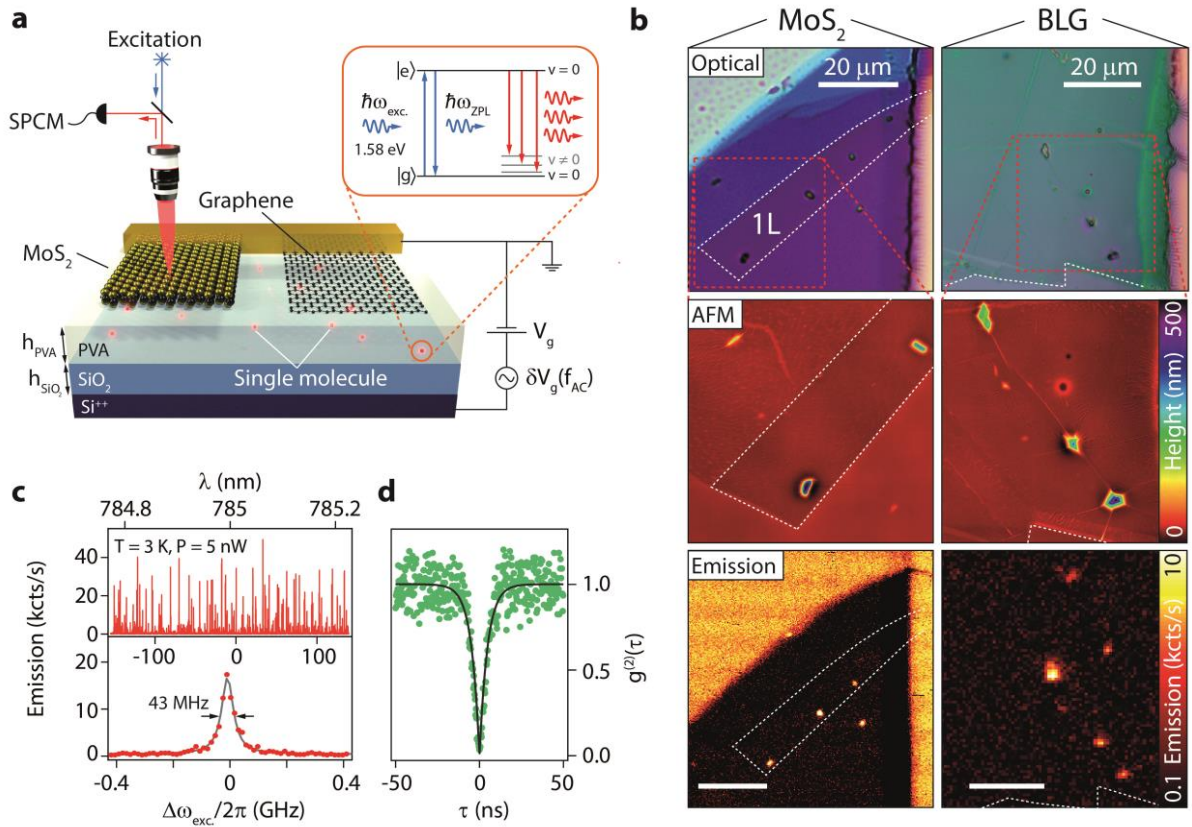

*Figure 1: Single molecules integrated with 2D materials.* **a** Hybrid device schematic; Atomically thin layers of graphene and MoS$_2$ cover fluorescent molecules embedded in a PVA film ($h_{PVA}$ = 300 nm) on SiO$_2$ ($h_{SiO2}$ = 285 nm). Single molecules are resonantly excited (inset) and their red-shifted fluorescence detected with a single photon counting module (SPCM). Electric fields are controlled by applying DC ($V_g$) and AC ($\delta V_g$) potentials to the Si$^{++}$ backgate. **b** Top to bottom: optical micrograph, AFM topography and DBT emission map for a MoS$_2$ (left column) and bilayer graphene (right column) device. White dashed lines outline the flakes. Scale bars are 20 μm. **c** Top panel: fluorescence excitation spectrum of an ensemble of single molecules in an uncovered nanocrystal at 3 K. Bottom panel: detail of two emission peaks with Lorentzian line shape (solid line), the narrower peak shows a FWHM of 43 $\pm$ 7 MHz. **d** Anti-bunching measurement for resonant excitation of a single peak as shown in **c**. The solid line is a fit to the data using a second-order correlation function.

A simplified DBT energy level scheme is shown in the inset of Fig. 1a. The transition of interest is the zero-phonon line (ZPL) between the ground vibrational levels (v=0) of the electronic ground and excited states. Upon resonant excitation of this transition, the molecule relaxes either to the electronic ground state (30-40% of emission[37]) or to a higher vibrational state (v≠0), thereby emitting a red-shifted photon. Scanning laser spectroscopy (see methods) on a pristine nanocrystal (Fig. 1c) at 3 K reveals a series of sharp peaks from a DBT ensemble. This peak dispersion arises from local variations of strain and charge and enables spectrally addressing a single molecule[38]. The narrowest peaks in such ensembles display a typical linewidth of $\Gamma = 2\pi \times (43 \pm 7)$ MHz.

To confirm that we can address single molecules in a nanocrystal, we measure statistics of photon emission in Hanbury-Brown and Twiss (HBT) configuration (Fig. 1d) and fit the normalised data with the second-order intensity autocorrelation function[39] $g^{(2)}(\tau) = 1 - C\, e^{-|\tau|/\tau_{DBT}}$, where $C$ is the contrast. At zero delay time, we find g$^{(2)}$(0) = 0.04 $\pm$ 0.02 < 0.5, which is signature of a single photon source. We extract an excited state lifetime[39] of $\tau_{DBT}$ = 4.7 $\pm$ 0.5 ns, implying a lifetime-limited linewidth $\Gamma_0 = \tau_{DBT}^{-1} = 2\pi \times (34 \pm 5)$ MHz, comparable to values observed in bulk anthracene crystals[37]. Therefore, single molecules in our device can exhibit lifetime-limited linewidth within the measurement error. In a solid state environment, this linewidth can be broadened by dephasing and near-field interactions, which can hamper the performance of emitters in nanostructured devices[36,40].

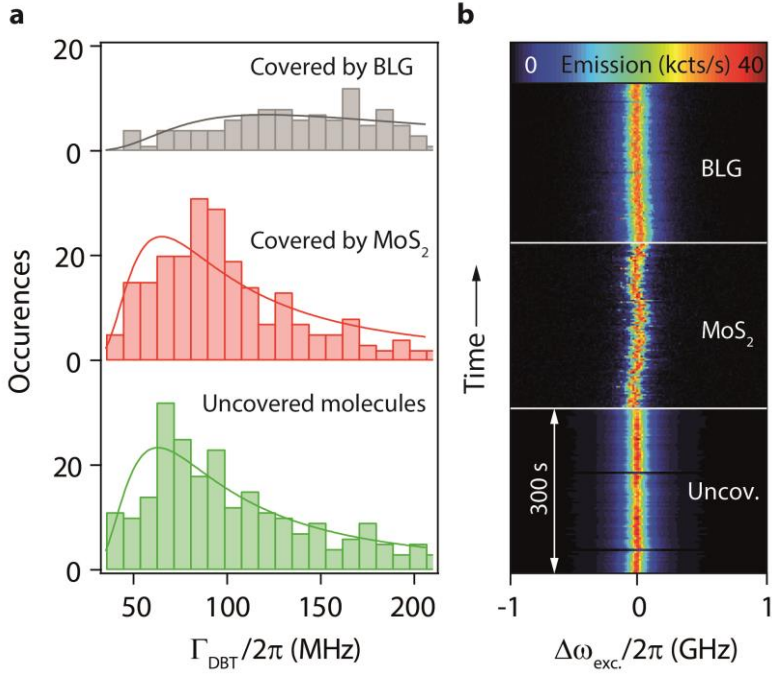

*Figure 2: Emission broadening and time stability in the presence of 2D materials. **a** Linewidth distribution for uncovered DBT (green), DBT covered by monolayer MoS$_2$ (red) and bilayer graphene (grey). Solid lines are fits to the data as described in the main text. **b** Time trace of single DBT molecule emission (3 s / line) for uncovered DBT (bottom) and DBT covered by monolayer MoS$_2$ (middle) and bilayer graphene (top).*

In our device, a 2D material is placed in the near-field of a single emitter (separation $d \ll \lambda$). In this regime, near-field interactions such as non-radiative energy transfer (nRET) from the molecule to the 2D material[41] and Casimir-Polder (CP) energy level shifts[8] are expected to occur. Both interactions scale divergently with separation as $d^{-4}$, in contrast to $d^{-3}$ for bulk interfaces[22]. The nRET process leads to a linewidth broadening while the CP energy level shift is a quantum effect related to the modification of vacuum fluctuations by the 2D interface. At fixed emitter-2D material separations, we can quantify the nRET contribution due to the presence of the 2D material by its impact on the emission linewidth. The CP contributions are difficult to quantify because for one specific emitter, the emission energy with and without the 2D material cannot be measured independently.

By measuring the emission linewidths of over 500 molecules (Fig. 2a) in three different configurations (uncovered, covered by graphene/MoS$_2$), we quantify the effect of proximity to 2D materials on emission linewidth. We quantify the most probable linewidth $\tilde{\Gamma}$ in each case by fitting the linewidth histogram with a Smirnov distribution[42] which describes single molecule linewidth distributions perturbed by long-range coupling to two-level fluctuators, e.g. in polymers[42]. For uncovered nanocrystals, we obtain $\tilde{\Gamma}_{uncov.} = 2\pi \times 62$ MHz, which is comparable to $\Gamma_0$. This implies that the anthracene nanocrystal is a highly stable and crystalline environment[35] and most molecules within it do not experience significant spectral diffusion due to defects or proximity to the surface[35]. While emitters covered by MoS$_2$ experience extremely weak spectral broadening ($\tilde{\Gamma}_{MoS_2} = 2\pi \times 65$ MHz), they are significantly broadened when covered by BLG ($\tilde{\Gamma}_{BLG} = 2\pi \times 108$ MHz) and their linewidth distributed over a larger range. We attribute this broadening to higher nRET efficiency to gapless graphene as electronic transitions can be optically excited over a large range of energies and in particular at $\hbar\omega_{ZPL}$. In contrast, MoS$_2$ has bandgap and excitonic resonances at energies > $\hbar\omega_{ZPL}$, resulting in weak nRET which preserves the narrow linewidth. These results show that neither the presence nor the integration process of 2D materials have a strong detrimental impact on the emission properties. In contrast, we find that the deposition of a commonly used transparent electrode material such as ITO results in the complete disappearance of DBT emission (see Supplementary Note 2) due to the deposition process. This highlights the weakly invasive nature of 2D material integration, which makes it particularly suitable for fragile quantum emitters.

To confirm that emission stability in time is preserved after 2D material integration, we measure DBT spectra over time for the three configurations (uncovered, covered by graphene/MoS$_2$) as shown in Fig. 2b. In each case, we observe stable emission intensity and fluctuations of ZPL frequency and linewidth below 42 MHz (see Supplementary Note 3). Slightly higher fluctuations of $\omega_{ZPL}$ - possibly induced by charge fluctuations - are resolved in the case of MoS$_2$. Overall, these results show that 2D semiconductors are particularly suitable for integration with sensitive quantum emitters at nanoscale proximity, introducing negligible perturbation in the photostability.

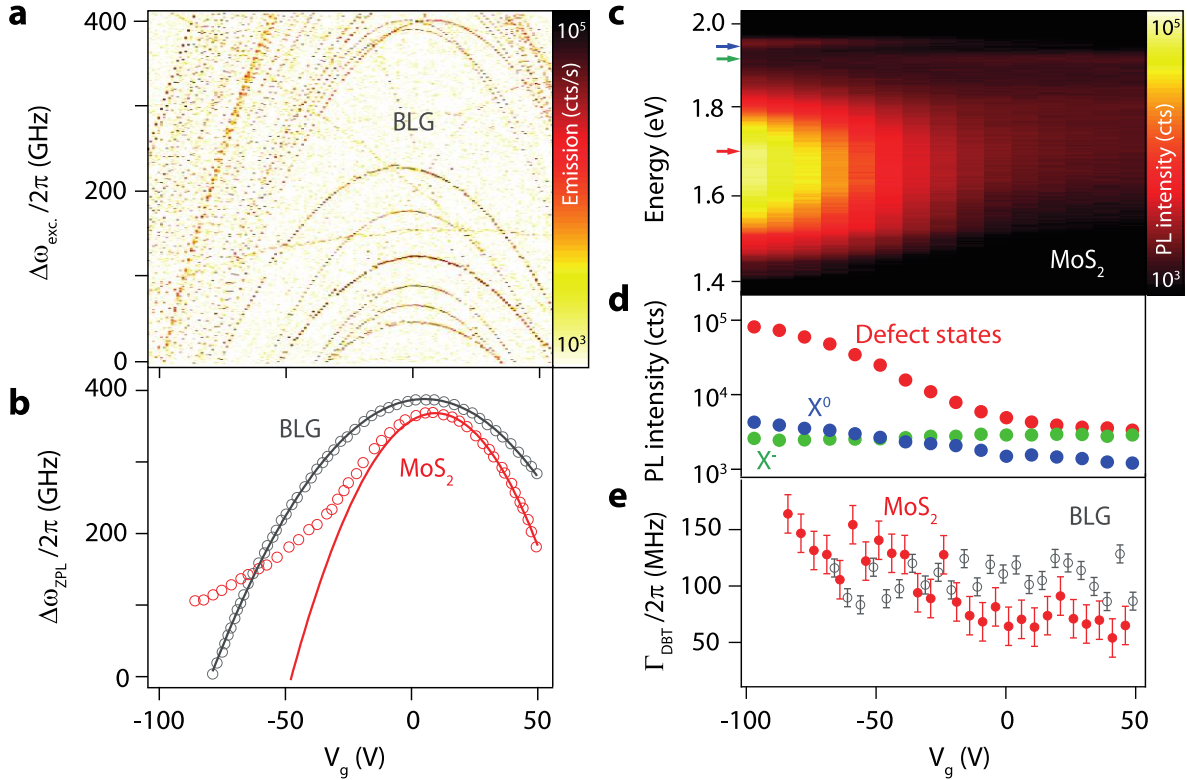

*Figure 3: Stark tuning of a single molecule with a 2D electrode.* **a** Spectral map showing the Stark shift of an ensemble of single molecules under BLG with back-gate voltage $V_g$. **b** Gate-induced line shift of a single molecule under BLG (grey circles) and monolayer MoS$_2$ (red circles). Solid lines are parabolic fits (see main text). **c** Back-gate dependence of MoS$_2$ photoluminescence (PL) spectra at 532 nm (2.33 eV) excitation. **d** PL intensity of exciton (1.96 eV, blue), negatively charged trion (1.93 eV, green) and defect band (1.7 eV, red) with $V_g$. **e** Emission linewidth for a single molecule under BLG (grey) and monolayer MoS$_2$ (red).

We now turn to electrical manipulation of single emitters. To achieve Stark[43] tuning of DBT emission energy, we apply an electric potential $V_g$ over the capacitor formed by the 2D electrode and the Si[++] back-gate, separated by a PVA/SiO$_2$ layer (see Fig. 1a). We reach comparatively large[44,45] field strengths above 2 MV.cm$^{-1}$ before dielectric breakdown takes place. DBT ensemble spectra below BLG at different $V_g$ (Fig. 3a) show a large, dominantly quadratic shift of the emitters in the whole ensemble on the order of hundreds of GHz ($\sim 10^4\, \Gamma_0/2\pi$), comparable to the inhomogeneous broadening of the entire ensemble[35]. While this tuning range is comparable to state-of–the-art devices employing semiconductor quantum dots[45] and diamond defects[46], we highlight that our device displays emission linewidths within a smaller footprint, which is attractive for nanophotonic integration. We also verify that DBT emission energy remains stable under large applied electric fields by repeating stability measurements as shown in Fig. 2b up to large values of $V_g$ (see Supplementary Note 3). We find small, gate-dependent emission energy drifts (< 3 MHz/s), likely due to small leakage currents in the device, which can be reduced by using a smaller 2D electrode area or by active feedback[29].

For a centrosymmetric molecule such as DBT, we expect a quadratic Stark shift with linear contributions arising from dipole moments induced by distortion of the molecule's insertion site[44]. The Stark shift is modelled as $\Delta\omega_{ZPL} = a|\vec{E}| + b|\vec{E}|^2$, where $a$ and $b$ are the linear and quadratic Stark coefficients, and functions of the dipole moment and polarizability change, respectively, between ground and excited state. Here, $\vec{E} = \vec{E}_{ext} - \vec{E}_0$ is the net electric field experienced by the molecule upon application of an external field $|\vec{E}_{ext}| = V_g/h_{tot}$, ($h_{tot} = h_{PVA} + h_{SiO_2} \sim 600$ nm) with a correction factor $\vec{E}_0$ that accounts for trapped charges and work function differences[47]. Fitting detuning data for different emitters, we extract typical Stark coefficients of $a \sim 500$ MHz/(kV cm$^{-1}$) and $b \sim -0.5$ MHz/(kV cm$^{-1}$)$^2$ (see Supplementary Note 4), comparable to reported results[34]. The observed quadratic stark shift shows that $E_{ext}$ is linear in $V_g$, as is expected for a metallic electrode such as BLG (Fig. 3b).

In contrast, emitters under MoS$_2$ deviate from the parabolic detuning below $V_{g,0} \sim -30$ V (Fig. 3b), where the detuning flattens and $\Delta\omega_{ZPL}$ is almost independent of $V_g$. We attribute this deviation to a gate-induced change of resistivity[48]: as the charge carrier density in a semiconductor changes more abruptly with $V_g$, we expect a sub-linear dependence of $E_{ext}(V_g)$ as excess charge carriers are depleted in MoS$_2$. Electrostatic doping of the MoS$_2$ electrode is independently confirmed by gate-dependent photoluminescence (Fig. 3c, d). The ratio of charged ($X^-$) to neutral ($X^0$) exciton PL intensities increases with $V_g$ due to n-doping of the MoS$_2$ electrode. We also observe a broad emission peak at lower energy for negative $V_g$ - associated to emission from defects in MoS$_2$[49] - which overlaps with $\hbar\omega_{ZPL}$. With the appearance of this defect band, we observe a linewidth broadening $\Gamma_{DBT}$ below $V_{g,0}$ for some emitters, which is absent in the case of a metallic BLG electrode (Fig. 3e). This suggests that defects could act as n-RET acceptors, leading to emission linewidth broadening of emitters close to MoS$_2$.

Our system combines large Stark tuning with lifetime-limited emitters that potentially enables adiabatic control up to a frequency set by their natural linewidth $\Gamma_0$. We investigate the range of this dynamical control by applying oscillating potentials $V_{tot} = V_g + \delta V_g(f_{AC})$ to monolayer graphene and MoS$_2$ electrodes. For fast modulation frequencies compared to the spectral acquisition time ($f_{AC} \gg t_{meas}^{-1} \sim 0.1$ Hz), DBT emission peaks show a splitting $\sigma(f_{AC})$ (Fig. 4a) proportional to the modulation amplitude $\delta V_g$ and the local Stark slope $\frac{\partial\omega_{ZPL}}{\partial V_g}$:

$$\sigma = \frac{\partial\omega_{ZPL}}{\partial V_g} \delta V_g \ T \quad (1)$$

Here, $T(f_{AC})$ is the transmission of the $RC$-low-pass filter governed by the effective device resistance $R$ which includes contact resistance as well as the sheet resistance of the 2D material and backgate capacitance $C$ (see Supplementary Note 5). At fixed $V_g$ and $f_{AC}$, the DBT emission spectra reflect the oscillation turning points of the modulating waveform (Fig. 4a). Under square modulation with a fixed amplitude $\delta V_g$, the splitting vanishes with increasing $f_{AC}$ for both MoS$_2$ and graphene electrodes (Fig. 4b), albeit at a lower frequency for MoS$_2$. From this measurement, we extract $T(f_{AC})$ (Eq. 1) and confirm low-pass behaviour with a characteristic cut-off frequency $f_{-3dB}$ (Fig. 4c). Interestingly, $T(f_{AC})$ shows a strong gate voltage $V_g$ dependence for MoS$_2$ electrode, which is not observed for graphene. This dependence is quantified by $f_{-3dB}$, which increases with $V_g$ and saturates at $\sim 5$ kHz $\ll \Gamma_0$ for $V_g > 0$ V (Fig. 4d). We attribute this behaviour to a change of sheet resistivity and contact resistance due to gate-induced electrostatic n-doping[48] of the MoS$_2$, consistent with PL measurements (Fig. 3c,d). As a result, $f_{-3dB} \propto (RC)^{-1}$ increases as we sweep $V_g$ from negative to positive values.

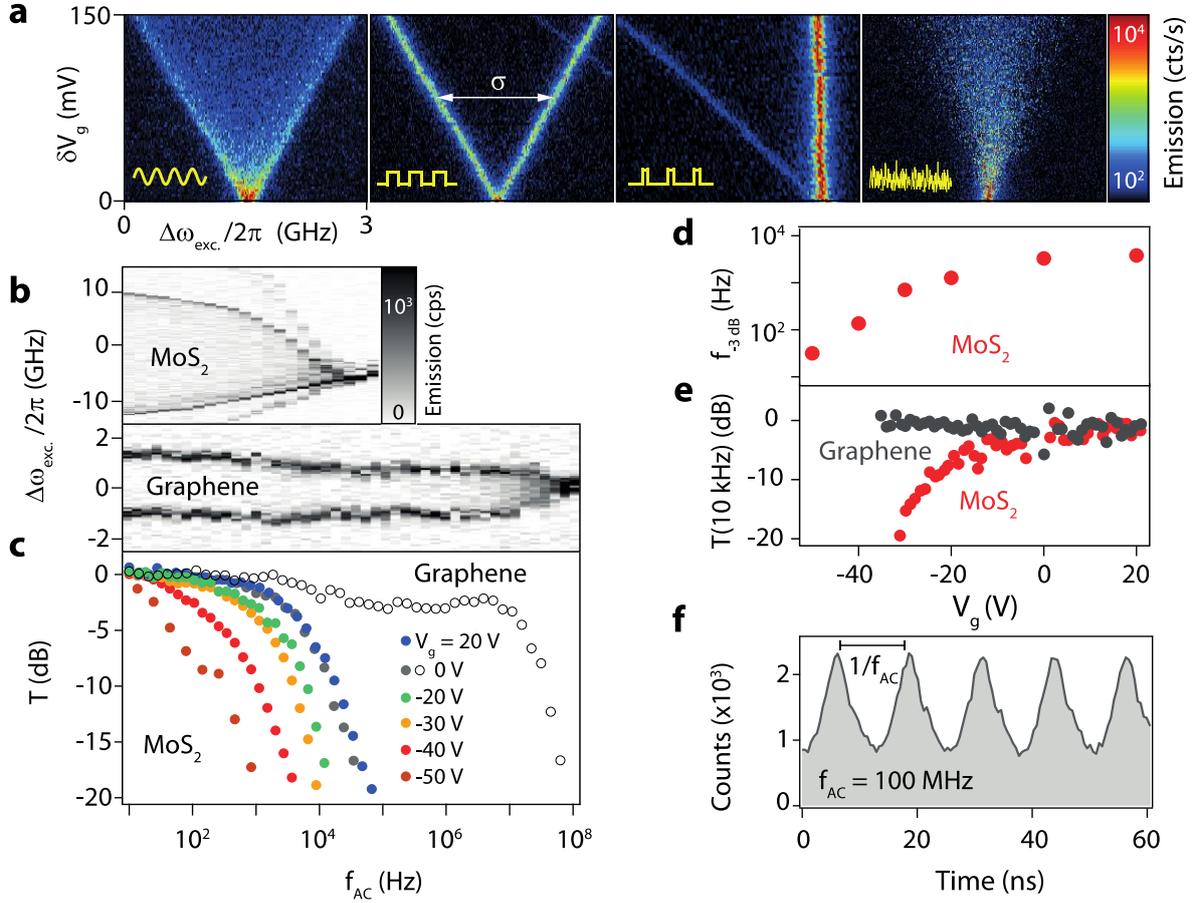

**Figure 4: Dynamical emission control of quantum emitter. a** Single molecule emission as a function of AC amplitude $\delta V_g$ for (left to right) sinusoidal, square and pulsed modulation ($f_{AC} = 1$ kHz), and for pink noise (100 kHz bandwidth). **b** Single DBT emission spectra vs. $f_{AC}$ ($V_g + \delta V_g = 20 + 1$ V) under MoS$_2$ (upper panel) and monolayer graphene (lower panel). **c** Transmission function $T(f_{AC})$ for a single molecule at different $V_g$ under MoS$_2$ (filled circles) and graphene (open circles). **d** Back-gate dependence of MoS$_2$ cut-off frequency $f_{-3dB}$. **e** Transmission $T(10\ kHz)$ as a function of $V_g$ for MoS$_2$ (red) and BLG (grey). **f** Histogram of time-resolved single molecule emission intensity modulation at $f_{AC} = 100$ MHz. The transition energy $\omega_{ZPL}$ is modulated electrically while the excitation energy $\omega_{exc}$ is kept constant, leading to a periodic emission strength in time.

To capture the gating efficiency of our 2D electrodes, we extract transmission $T$ at fixed $f_{AC} = 10$ kHz varying $V_g$ (Fig. 4e, and SI). The strong reduction of $T$ for $V_g < 0\ V$ is consistent with a large gate-induced change of resistance in the semiconducting MoS$_2$ device, while graphene's metallicity maintains $T \sim 1$ over the full $V_g$ range. Therefore, graphene enables Stark tuning of emitters over a large energy range and at high frequency at the expense of weak linewidth broadening. To extend this modulation bandwidth, we reduce the gate capacitance $C$ by using a thicker PVA dielectric ($h_{PVA} = 800$ nm). Then, we modulate $\hbar\omega_{ZPL}$ around a fixed excitation energy by applying $\delta V_g$ at $f_{AC} = 100$ MHz $\sim \Gamma_0/2\pi$ to a graphene electrode. Using a time-correlated single photon counter synchronised to the modulation $\delta V_g$, we observe a periodic oscillation of emission intensity (Fig. 4f) with a modulation depth of $\sim 50\ \%$. This oscillation is a signature of the excitation laser periodically exploring a fraction of the emitter's absorption line. These measurements show that the dynamical modulation bandwidth of our devices - determined by the 2D electrode material and device geometry – approaches $\Gamma_0$. Conversely, single emitters act as local nanoprobes of the 2D material's electronic properties.

In conclusion, we demonstrate a novel hybrid device where 2D materials are integrated with lifetime-limited single photon emitters to achieve broadband and fast emission energy tuning. Our results highlight the potential of atomically thin electrodes for integration with sensitive quantum emitters in a nanoscale device without perturbing narrow emission linewidth, in contrast to commonly employed bulk transparent electrodes. At high frequencies, Stark modulation reveals low-pass transmission behaviour related to the 2D materials' sheet resistivity. Using a graphene electrode, we show emission energy modulation at frequencies approaching the emitters' linewidth. Our device thus provides resonant tuning and high-frequency modulation of SPS on chip, required for obtaining indistinguishable and synchronised single photons. Finally, our device is a platform for studying novel forms of light-matter interaction with plasmon-polaritons in graphene[9] and exciton-polaritons in TMDs[10] at the single excitation level. Conversely, strong coupling to such excitations could allow the observation of normally forbidden higher-order transitions of the emitter[50].

**Acknowledgments:**

We acknowledge Rinu Abraham Maniyara for advice regarding transparent electrodes. F.H.L.K. acknowledges financial support from the Government of Catalonia trough the SGR grant (2014-SGR-1535), and from the Spanish Ministry of Economy and Competitiveness, through the "Severo Ochoa" Programme for Centres of Excellence in R&D (SEV-2015-0522), support by Fundacio Cellex Barcelona, CERCA Programme / Generalitat de Catalunya and the Mineco grants Ramón y Cajal (RYC-2012-12281) and Plan Nacional (FIS2013-47161-P and FIS2014-59639-JIN). Furthermore, the research leading to these results has received funding from the European Union Seventh Framework Programme under grant agreement Orquid.

**Methods:**

**Device fabrication.** We make a suspension of anthracene nanocrystals hosting DBT molecules in PVA by reprecipitation. The suspension is spin-cast onto a p-doped Si wafer coated with 285 nm thermally grown $SiO_2$. Electrodes are deposited onto the coated chip by thermal evaporation of 100 nm Au through a shadow mask. Nanocrystals close to the electrode displaying DBT fluorescence are localised at room temperature using off-resonant wide-field illumination. Bulk $MoS_2$ and graphite is exfoliated mechanically using commercial polydimethylsiloxane (PDMS) sheets. Mono- and bilayer flakes of both materials are identified optically by absorption contrast (see Supplementary Note 6) and transferred onto anthracene nanocrystals in PVA by dry stamp transfer.

**Electrical device actuation and optical readout.** Measurements are performed under vacuum in a cryostat at 3 K. Our hybrid devices are actuated electrically using a low-noise voltage source and an arbitrary waveform generator to provide DC and AC voltages. Using a custom-built confocal microscope to locally illuminate the device, we excite single molecules with a tuneable 785 nm laser with circular polarisation at 5 nW, well below saturation power ~ 20 nW (see Supplementary Note 7). Red-shifted single photon emission is detected with a single photon counting module (SPCM) combined with spectral ZPL filtering by a long-pass filter. Emission maps are made using pseudo-broadband excitation of DBT ensemble fluorescence by fast modulation (200 Hz) of the laser detuning compared to the SPCM integration time (~10 ms). $MoS_2$ photoluminescence is excited using a 532 nm laser and detected with a spectrometer.

**Author contributions**

K.G.S., A.R.-P., C.T. and F.H.L.K. conceived the experiment. K. G. S. and C.C. made the samples with the support of S.P., P.L. and C.T. Measurements and analysis were performed by K.G.S., C.C. and A.R.-P. K.G.S, A.R.-P. and F.H.L.K. wrote the manuscript with critical comments from all authors.